\documentclass[aps,prb,showpacs,preprint,amssymb]{revtex4}
\usepackage[T1]{fontenc}
\usepackage{amsmath}
\usepackage{graphicx}
\usepackage{bm}
\usepackage[cp1250]{inputenc}
\newcommand{\be}{\begin{equation}}                  
\newcommand{\ee}{\end{equation}}                    
\newcommand{\bml}{\begin{multiline}}                
\newcommand{\eml}{\end{multiline}}                  
\newcommand{\bgt}{\begin{gather}}                    
\newcommand{\egt}{\end{gather}}
\newcommand{\ba}{\begin{eqnarray}}                  
\newcommand{\ea}{\end{eqnarray}}
\newcommand{\bal}{\begin{align}}                    
\newcommand{\eal}{\end{align}}
\newcommand{\pt}{\partial}
\newcommand{\di}{\mathrm{d}}

\newcommand{\PDA}{P^{\scriptscriptstyle\mrm{DA}}}

\newcommand{\cF}{\mathcal{F}}

\newcommand{\cJ}{\mathcal{J}}
\newcommand{\cD}{\mathcal{D}}

\newcommand{\cM}{\mathcal{M}}
\newcommand{\bk}{{\bm k}}
\newcommand{\bq}{{\bm q}}
\newcommand{\bu}{{\bm u}}
\newcommand{\bv}{{\bm v}}
\newcommand{\vep}{{\varepsilon}}
\newcommand{\veps}{{\varepsilon}_s}
\newcommand{\vepF}{\varepsilon_{\scriptscriptstyle\mrm{F}}}

\newcommand{\vepq}{\varepsilon_{\bq}}

\newcommand{\vepk}{\varepsilon_{\bk}}
\newcommand{\vepkq}{\varepsilon_{\bk+\bq}}
\newcommand{\jex}{j_{\mrm{ex}}}
\newcommand{\zq}{z_{q}}
\newcommand{\kB}{{k_{\scriptscriptstyle\mrm{B}}}}
\newcommand{\kBT}{{k_{\scriptscriptstyle\mrm{B}}T}}
\newcommand{\kF}{{k_{\scriptscriptstyle\mrm{F}}}}
\newcommand{\qD}{{q_{\scriptscriptstyle\mrm{D}}}}
\newcommand{\vF}{{v_{\scriptscriptstyle\mrm{F}}}}
\newcommand{\cG}{\mathcal{G}}
\newcommand{\cH}{\mathcal{H}}
\newcommand{\Cph}{C^\mathrm{ph}}
\newcommand{\Sph}{S^\mathrm{ph}}
\newcommand{\Pph}{P^\mathrm{ph}}
\newcommand{\Pmag}{P^\mathrm{mag}}
\newcommand{\Cmag}{C^\mathrm{mag}}
\newcommand{\Smag}{S^\mathrm{mag}}
\newcommand{\Wmag}{W^\mathrm{mag}}
\newcommand{\Wph}{W^\mathrm{ph}}
\newcommand{\rhomg}{\rho^\mathrm{mag}}
\newcommand{\rhoph}{\rho^\mathrm{ph}}
\newcommand{\TD}{T_{\!\mathrm{D}}} 
\newcommand{\TC}{T_{\!\mathrm{C}}} 
\newcommand{\cS}{\mathcal{S}}
\newcommand{\cSph}{\cS^{\mrm{ph}}}
\newcommand{\cSmg}{\cS^{\mrm{mag}}}

\newcommand{\mrm}{\mathrm}
\newcommand{\ovl}{\overline}
\newcommand{\THeta}{\mathit{\Theta}}
\newcommand{\DElta}{\mathit{\Delta}}
\newcommand{\XI}{\mathit{\Xi}}
\begin{document}

\title{Comments on the  thermoelectric power\\ of the 
f-electron metallic compounds}
\author{A.E.\,Szukiel}

\affiliation{Institute of Low Temperature and Structure Research, Polish Academy of Sciences, Wroc\l aw, Poland}

\date{\today}

\begin{abstract}

The anomalous temperature variation of the thermoelectric power in the f-electron metallic compounds, namely the sign reversal or the maxima, is sometimes interpreted as resulting from the conduction electrons scattering in Born approximation on the acoustic phonons and  on the localized spins in the s--f exchange interaction.
The experimenters rely on the results of some theoretical works where such thermoelectric power behavior was obtained within these simple models.
In the present paper we prove that  neither the electron--phonon scattering nor the magnetic s--f scattering in the Born approximation (nor both of them)  do lead to the effects mentioned above.

\end{abstract}

\maketitle

\section{\label{sec:Introduction}Introduction}

Over the last decades  can be seen growing interest in the thermoelectric power of  f-electron metallic compounds.
The application studies concern mainly the strongly correlated electron systems (SCES), which seem to be promising as the thermoelectric materials.\cite{pa,hvar}
The fundamental researches are focused both on the SCES, see e.g. Refs\,[\onlinecite{a1,a2,a3,a4,a5,a6,a7,a8,a9}] as well as on the well-localized f-electron systems (WLS) -- see  Refs\,[\onlinecite{n0,n1,n2,n3,n4,n5,n6,n7}].
The thermoelectric power (TEP) of the SCES reaches typically the values one or two orders greater than those of the WLS,  but in both  material groups it exhibits the anomalies as the sign change and/or the maxima (minima). In SCES these anomalies occur, in general,  at the temperatures much higher than in WLS.  The main reasons of such TEP behavior in magnetic metals are associated with phonons and the "magnetic" electrons of incomplete  shell. If we disregard  the  anomalies caused by the phonon-drag or the magnon-drag (see e.g.  Ref\,[\onlinecite{leg,bla1}]), we have to deal with those due to the conduction electron scattering by phonons  and  those  caused by the  conduction electron and f-shell interaction (k-f interaction).   
In SCES the dominant role plays the k-f interaction originating from the conduction and f-electron hybridization. 
  Theoretical models that help to explain 
 the great values and the high-temperature maxima (minima) with the possible sign change of TEP in SCES systems are founded on this type of interaction, see e.g. Refs\,[\onlinecite{t1,t2,t3,t4,t5}].

 In WLS even  more important role than hybridization can play the Coulomb interaction\,\cite{ful}. For 4f electron systems both type of  interaction were described by Hirst (1978) within the same   form of generalized k-f interaction ( discussed briefly in Ref\,[\onlinecite{ful}]). The contribution to the TEP anomalies from that interaction for  rare-earths paramagnetic systems was considered in Ref\,[\onlinecite{tak}]. It has been  shown that the anomaly arises in the third order of the scattering interaction when it fulfills some symmetry condition with respect to the symmetries of the ground and the excited f-electron states in the crystalline field. In particular the anomaly can be caused by the aspherical Coulomb interaction, whereas the isotropic s-f exchange interaction is excluded regardless the character of the crystal-field splitting.\cite{tak}. The anomaly manifests itself at  the temperature corresponding to the half of the excited state energy as the maximum, or as the minimum combined with the sign reversal. Both these cases were illustrated in Ref\,[\onlinecite{sier}] for  two examples of rare-earth intermetallics,  and very good agreement between  theory and  experiment was achieved.
In the ordered f-electron systems also the isotropic s-f exchange scattering may contribute to the anomalies of thermoelectric power. In ferromagnets the conduction-electron band splitting due to this interaction causes an asymmetry in the scattering intensities for the spin-up and spin-down electrons in their scattering on the system of the f-electron localized moments.   This leads to the maxima of TEP  below the Curie temperature, as it was shown in the Born approximation in Ref\,[\onlinecite{kas}].
The effect occurs both when the f-electron moments are described in the molecular field approximation (MFA) and when the spin-wave approximation is applied.

However the considerations in  Ref\,[\onlinecite{kas}] do not include the influence of the crystal field on the f-electron level (in the single-ion approximation)  nor on the collective excitations in f-electron system. Thus the results of this work are applicable only for  ferromagnets without the crystal field splitting (see  e.g.  Ref\,[\onlinecite{pin}]).
Another possible cause of anomalous TEP at low temperatures may be  the conduction electron scattering on phonons, considered in  the second Born approximation. This effect was demonstrated  in Ref\,[\onlinecite{niel}], and was called there  "the phony phonon-drag", because  it gives the TEP maxima of the similar magnitude and the temperatures of occurrence as those attributed previously to phonon-drag.

We have presented above some known mechanisms of the thermoelectric power anomalies
 possible for the f-electron metallic compounds.  Among these  referring to the low-temperature anomalies, Refs\,[\onlinecite{tak, kas, niel}],  the  first two are related to  WLS and the last may concern also SCES.

 It should be noted that  all of these mechanisms (including those characteristic for SCES\cite{t1,t2,t3,t4,t5} ) rely on  the non-standard models: the second Born approximation for the scattering probability or the two-band model for the conduction electron experiencing the s-f interaction.

In some experimental works (see e.g.  Refs\,[\onlinecite{a2,a3,a4,a5,a6,a7,a8,a9,n0,n1,n2,n3,n4,n5,n6,n7}])
one can find the interpretations of TEP anomalies  as caused by  the isotropic s--f exchange scattering (in the one-band approach for the conduction electrons) and/or the electron--phonon scattering -- considered in the Born approximation.
These interpretations  are based on the results of  Refs\,[\onlinecite{aus,aus1,aus2}] where the TEP maxima and the sign reversal were obtained within these oversimplified models. 

The authors of Refs\,[\onlinecite{aus,aus1,aus2}] performed the numerical calculations of the electron--phonon contribution to the thermoelectric power, $\cSph(T)$,  according to the formula derived 
within the variational method of solving the Boltzmann equation.\cite{koh,koh1,zim}
However, in their calculations they have not applied the scattering matrix elements presented in Refs\,[\onlinecite{koh,koh1,zim}], but other, received by themselves.\cite{aus, aus1,aus2} \
As the result, they have obtained $\cSph(T)$ exhibiting the sign reversal and maxima.
Calculating additionally the contribution from the s--f scattering in Born approximation (but neglecting the conduction band splitting), and next the total thermoelectric power $\cS(T)$ with the use of the Matthiessen's rule for the scattering matrix elements, they obtained the same effects.\cite{aus}
In consequence, they could reproduce the TEP maxima occurring in some rare earths intermetallic systems like $RE$Al$_2$, Ref.\,[\onlinecite{gra}].

Notwithstanding, as we show in the present paper,
neither the electron--phonon scattering itself, nor in the combination with the isotropic s--f scattering (within the model of scattering applied in Refs\,[\onlinecite{aus}]) do lead to the TEP sign reversal nor its maxima.
We prove this by calculating and analyzing the total thermoelectric power $\cS(T)$ and the contributions from the s--f scattering, $\cSmg(T)$, and from the electron--phonon scattering, $\cSph(T)$,  with the use of the same formula for the TEP as in  Refs\,[\onlinecite{aus,aus1,aus2}].
We also discuss the variation of $\cSph(T)$ and $\cS(T)$ in dependence on material parameters for the same parameter values as used in the cited works.
In conclusion we indicate the crucial points of  calculations reported in Refs\,[\onlinecite{aus,aus1,aus2}] that led to
 different results.

\section{\label{sec:The model and the method of calculation}{The model and the method of calculation}}

We consider the thermoelectric power $\cS(T)$ of cubic ferromagnetic metal with localized spins, assuming, similarly as in Ref.\,[ \onlinecite{aus}], that it results from the free electron scattering in Born approximation on acoustic phonons in the Debye model and on the localized spins in the mean field approximation (MFA).
For the calculation of $\cS(T)$ and its particular phonon $\cSph(T)$ and magnetic $\cSmg(T)$ contributions, we use the formula derived by Kohler, Refs\,[\onlinecite{koh,koh1}], through the variational method of solving the linearized Boltzmann equation.
The essence of the method is the statement, known as the variational principle, that the solution of Boltzmann equation (the wave vector $\bk$-dependent non-equilibrium electronic distribution function) should realize the maximum of the functional describing the production of the entropy induced by the scattering in the steady state (for transport process).
In this way, solving Boltzmann equation becomes equivalent to the variational problem and has been solved in Ref.\,[\onlinecite{koh}] with the use of Ritz method.
After expansion of the searched solution with respect to some base $\phi_i(\bk)$, $i=1,...,n$, the coefficients of the expansion are found from the system of linear equations following from the variational principle.
With the use of the found coefficients, the electrical $\bm{J}$ and the thermal $\bm{U}$ currents can be represented as the combinations of the trial currents $J_i$\,, $U_i$\,,
\ba\label{44} 
J_i  &=&  -e \int\!\di\bk \,\, \frac{-\partial f^0_{\bk}}{\partial\vepk} \,\,
        \phi_i(\bk) (\bv_{\bk} \cdot \bu),            \nonumber\\
U_i  &=&  - \int\!\di\bk \,\, \frac{ -\partial f^0_{\bk}}{\partial\vepk} \,\,
        \phi_i(\bk) (\bv_{\bk} \cdot \bu) (\vepk -\zeta(T)),
\ea
where
$f^0_\bk =f^0(\vepk) = (\exp[(\vep_\bk-\zeta(T))/\kBT]+1)^{-1}$
is the equilibrium electron distribution function, $\bu$ denotes the external field direction, $\bv_{\bk}$ -- electron velocity,
$\vepk = (\hbar\bk)^2/2m$ -- electron energy,  $\zeta(T)$ -- the chemical potential.

Substituting currents
$\bm J$, $\bm U$
to the Onsager transport equations one gets the transport coefficients expressed by the products of the trial currents $J_i$\,, $U_i$ and of the elements of the $n$-dimensional scattering matrix $P_{ij}$
\be\label{2}
\begin{split}
   P_{ij}(T)  =  \frac{V}{\kBT}
   \int\!\di\bk  \int\!\di\bk'\,  C(\bk,\!\bk')\,\, f^0_\bk (1-f^0_{\bk'})\,\, u_{ij}(\bk,\!\bk')       \\
    u_{ij}(\bk,\bk')  =  [\phi_i(\bk)-\phi_i(\bk')] [\phi_j(\bk)-\phi_j(\bk')].
\end{split}
\ee
Here
$C(\bk,\bk')$ is---for a given scattering process---the transition probability per unit time for the free electron scattered from the state $\bk$ to the state $\bk'$.

 The expressions for the electrical  and the thermal  conductivity, and for the thermoelectric power  of metals (for the highly degenerated electron gas) in the variational approximation of the $n$-th order were derived in Ref.[\onlinecite{koh}] with the use of the base functions
\be\label{1} 
\phi_i(\bk) = (\bk\cdot\bu) (\vep_\bk-\zeta(T))^{i-1}, \qquad i=1,\ldots,n\,.
\ee

The lowest-order approximation for the thermoelectric power was obtained with $\phi_1(\bk)$, $\phi_2(\bk)$ base functions, as it is seen from the formula in Eq.\,(27), Ref.\,[\onlinecite{koh}].
The formula (known also as the Ziman's formula) and its detailed derivation can also be find in Ziman's monograph,\cite{zim} see Eq.\,(9.12.13) therein.
It can be written in the form
\be\label{3} 
\begin{split}
         \cS(T) =&\ \frac{\pi^2}3 \frac{\kB}{e}\, \frac{\ovl{S(T)}}{\ovl{P_{22}}}\,,                   \\
     \ovl{S(T)} =&\ \ovl{P_{12}} - \frac32 \, \frac{\kBT}{\vepF}\, \ovl{P_{22}} - \frac{\pi^2}2 \frac{\kBT}{\vepF}\, \ovl{P_{11}},
\end{split}\ee

  after using  the reduced  $\ovl{P_{ij}}=P_{ij}/(\kBT)^{i+j}$ form of the scattering matrix elements and assuming   that $e>0$ ( the electron charge is $-e$).

The lowest order variational approximations for the electrical $\rho$  and the thermal $W$ resistivity rely on the one-dimensional sets of the base functions
-- $\phi_1(\bk)$ in the first case and
$\phi_2(\bk)$ in the second one\cite{koh}
\ba\label{4} 
\rho(T)  &=& \frac{ \ovl{P_{11}} }{ J_1^2 }                                            \nonumber\\
 W(T)    &=&   T\, (\kBT)^2\, \frac{ \ovl{P_{22}} }{ U_2^2 }\,,
\ea
where $J_1$\,, $U_2$ are the trial currents (\ref{44}), compare Eq.\,(19a) and the first of Eqs.\,(20a) in Ref.\,[\onlinecite{koh}].

Considering the phonon $\cSph(T)$ and the magnetic $\Smag(T)$ contributions to the thermoelectric power,
we will use the formula (\ref{3}) defined correspondingly with the electron--phonon $\Cph(\bk,\bk')$
and the s--f (magnetic) $\Cmag(\bk,\bk')$ scattering probability.
The total thermoelectric power $S(T)$ (\ref{3}) is calculated with
$C(\bk,\bk')= \Cph(\bk,\bk') + \Cmag(\bk,\bk')$, according to Matthiessen rule.
The same rule concerns the scattering matrix elements (\ref{2}) and, consequently, the electrical and the thermal resistivity (\ref{4})
\be\label{6} 
\begin{split}
\rho(T)=&\ \rhomg(T) + \rhoph(T), \\
W(T) =&\ \Wmag(T) + \Wph(T)
\end{split}
\ee

From (\ref{3})--(\ref{6}) the Kohlers rule follows
\be\label{8}
S(T) = \frac{ \Sph(T)\,\Wph(T) + \Smag(T)\,\Wmag(T) }{W(T)}\,,
\ee
which in its original form concerns the case when the total thermoelectric power was the result of the electron--phonon and the electron--impurity scattering,  see Eq.\,(8) in Ref.\,[{\onlinecite{koh1}].

\section{\label{sec:Magnetic Contribution}
Magnetic contribution to the thermoelectric power:\\ \protect{\hspace*{2em}} electron scattering on magnetic ions}


$\Cmag(\bk',\bk)$, the transition probability for the conduction electron experiencing  the s-f exchange interaction, can be expressed, after summation with respect to all possible changes of the electron spin, by
$\Im\mrm{Tr} \chi(\bq,\hbar\omega) / (\exp[\hbar\omega/\kBT]-1)$,
for
$\bq=\bk'-\bk$ and $\hbar\omega= \varepsilon'-\varepsilon$,
where $\chi(\bq,\hbar\omega)$ is the susceptibility function for the system of localized f-electrons, and
$\vep' = (\hbar\,\bk')^2/2m$, $\vep = (\hbar\,\bk)^2/2m$, see e.g. Ref.\,[\onlinecite{hes1}].
Using the MFA susceptibilities for the cubic ferromagnet\cite{jen} one gets\\
\ba\label{0}
\Cmag(\bk',\bk) &=& \frac{2\pi\jex^2\,(g\!-\!1)^2N}{\hbar}
\left[ \left( \frac{\langle J^z \rangle \delta(\vep'\!-\!\vep\!-\!\DElta)}{\exp[\DElta/\kBT]\!-\!1}
 + \frac{\langle J^z \rangle \delta(\vep'\!-\!\vep\!+\!\DElta)}{1\!-\!\exp[-\DElta/\kBT]} \right)
 + (\delta J^z)^2 \delta (\vep'\!-\!\vep) \right]                                                                \nonumber\\
(\delta J^z)^2 &=& \langle(J^z)^2\rangle - \langle J^z \rangle^2 = J(J+1)
- \frac{\langle J^z \rangle}{\tanh(\DElta/(2\kBT))} - \langle J^z \rangle^2,
\ea
where $\jex$ denotes the energy of the s--f exchange interaction,
$N$ is the number of ions per unit volume,
$\langle J^z \rangle$ is the MFA thermodynamical expectation value of the $z$-component of the f-electron total angular momentum;
$\DElta = 3\kBT_{\!c} \langle J^z \rangle / J(J\!+\!1)$ is the molecular field energy,
$\TC$ the Curie temperature and $J$ is the maximal eigenvalue of the operator $J^z$.

The components in round brackets refer to the inelastic scattering -- the first describes the scattering with the energy absorption
and the second with the emission.
The last component in square brackets refers to the elastic scattering.

Since $\Cmag(\bk,\bk')$ is the even function with $\bk$ and with $\bk'$
\be\label{02}
\Cmag (\bk',\bk) =  \Cmag (-\bk',\bk) = \Cmag (\bk',-\bk),
\ee
and the base functions $\phi_i(\bk)$ (\ref{1}) are odd,
the only components of $u_{ij}(\bk,\bk')$ giving contribution to the scattering probabilities
$\Pmag_{ij}(T)$ (\ref{2}) are the products $\phi_i(\bk)\,\phi_j(\bk)$, $\phi_i(\bk')\,\phi_j(\bk')$.
For this reason $\Pmag_{ij}(T)$ can be expressed in the form
\be\label{03}
\Pmag_{ij}(T)= \frac{1}{4\pi^3}
\int \di\bk \left( -\frac{\pt f^0(\vepk)}{\pt \vepk}\right)\frac{1}{\tau(\vep_{\bk})} \phi_i(\bk)\,\phi_j(\bk),
\ee
where  $\tau(\vep_{\bk})$ is the relaxation time depending only on the electron energy
\be\label{001}
\frac1{\tau(\vep_{\bk})} = \frac1{4\pi^3}
\int \!\di {\bk'} \Cmag (\bk,\bk') \frac{1-f^0(\vep_{\bk'})}{1-f^0(\vep_{\bk})},
\ee
and the identity
$ f^0(\vepk)(1-f^0(\vep_{\bk'})) \equiv \kBT\,\left( -\pt f(\vepk)/\pt \vepk\right) (1-f^0(\vep_{\bk'}))/(1-f^0(\vep_{\bk}))$ was applied.

Performing the integration with respect to $\bk$ and $\bk'$  in the standard way (see e.g. Ref.\,(\onlinecite{zim}) one gets
\be\label{50d}
\Pmag_{ij}(T) = \frac{2m}{3\hbar^2}
\int \limits_0^\infty \!\di\vep\,\cD(\vep)\,\vep(\vep-\vepF)^{i+j-2}
\left(-\frac{\pt f^0(\vep)}{\pt\vep} \right)  \frac1{\tau(\vep)},
\ee
where
$\cD(\vep) = (2m)^{3/2}\,\vep^{1/2} / (2\pi^2\,\hbar^3)$
is the density of states for the free conduction electrons, and
\ba\label{50e}
&& \frac1{\tau(\vep)} = \frac{\pi \jex^2 N}{\hbar} \,\cD(\vep) \left(\cM(y(\vep)) + (\delta\,J^{z})^2\right),       \nonumber\\[1ex]
     \cM(y(\vep))  &=& \langle J^z \rangle \left(\frac{1+\exp[-y]}{(\exp[z]-1)(1+\exp[-(y+z)])}
                    +  \frac{1+\exp[-y]}{1-(\exp[-z])(1+\exp[-(y-z)])} \right)                                     \nonumber\\[1.5ex]
&&                y  = (\vep-\vepF)/\kBT, \qquad  z=\Delta/\kBT.
\ea

Next we will calculate (\ref{50d}) with the use of the Sommerfeld expansion (\ref{22b}) confining ourselves to the first non-vanishing term in the approximation of the strong degeneration of the electron gas.
Noting, additionally, that $\cM(y)$ is even, we get $\pt\cM(\vep)/ \pt\vep\mid_{\vep=\vepF} = 0$, and in consequence $\Pmag_{ij}$ in the form
\ba\label{50f}
\Pmag_{11} = \Pmag_{11} \left.(\vep)\right|_{\vep=\vepF}
           &=& \frac{(2m)^{5/2}}{6\pi^2\hbar^5}\,\vep^{3/2}\left. \frac{1}{\tau(\vep)} \right|_{\vep=\vepF}     \nonumber\\[1ex]
\Pmag_{12} = \frac{\pi^2}3 (\kBT)^2\;\left.\frac{\pt\Pmag_{11}(\vep)}{\pt\vep}\right|_{\vep=\vepF}
           &=& \frac{2\pi^2}3 \frac{(\kBT)^2}{\vepF}  \Pmag_{11}                                                 \nonumber\\[1ex]
\Pmag_{22} &=& \frac{\pi^2}3 (\kBT)^2 \left. \Pmag_{11}(\vep)\right|_{\vep=\vepF}.
\ea

The magnetic contribution to the thermoelectric power we get substituting in (\ref{3}) $\ovl{\Pmag_{ij}} = \Pmag_{ij} / (\kBT)^{i+j-2}$  in place of $\ovl{P_{ij}}$.
After  some algebra we can write it in the form
\be\label{50h}
\cSmg(T) = - \frac{\pi^2}3\, \frac{\kB}{e}\, \frac{\kBT}{\vepF}.
\ee

This result is the same as that obtained from Mott formula when the solution of Boltzmann equation is described by relaxation time depending on the electron energy as $\tau(\vep)\sim \vep^{-1/2}$.
The exemplification of that case is the elastic electron scattering on the ionized impurities, see e.g. Ref.\,[\onlinecite{bla1}].

For the magnetic part of the electrical $\rhomg(T)$ and the thermal $\Wmag(T)$ resistivities
we substitute in (\ref{4}) $\ovl{\Pmag_{ij}}$ with  $1/\tau(\vepF)$
\be\label{50k}
\frac1{\tau(\vepF)}  =  \frac{G^2m\kF N}{\pi\hbar^3}
                   \left[J(J+1) - \langle J^z \rangle^2 - \langle J^z \rangle \tanh(\DElta/2\kBT) \right],
\ee
and  $J_1$,  $U_2$  in the form
\be\label{500k}
J_1  =  \frac{e\,\kF^3}{(3\pi^2\,\hbar)},   \qquad    U_2 = \frac{J_1\,\pi^2(\kBT)^2 }{3e} ,
\ee
resulting from the integration in (\ref{44}) with respect to $\bk$ in the manner described above.
The final results are as follows
\ba\label{50m}
\rhomg(T) &=& \rhomg_0 \left[ J(J+1) - \langle J^z \rangle^2 - \langle J^z \rangle \tanh(\DElta/2\kBT) \right] \nonumber\\
\Wmag(T) &=& \frac{\rhomg(T)}{L_0\,T},\,
\ea
where $\rhomg_0 = 3\pi^2\jex^2 m N / (2e^2\hbar\vepF)$,
and $L_0 = \pi^2\,\kB^2 / (3e^2)$ is the Lorentz number.

For $T\ll \TC$,  after approximating
$\langle J^z \rangle_{\mrm{MFA}} \sim J-\exp[-3\TC/((J\!+\!1)\,T) ]$, Ref.[\onlinecite{jen}],
there is
\be\label{51a}
\rhomg  \sim  \rhomg_0\,2J \exp[-3\TC / ((J+1)\,T)],
\qquad
\Wmag \sim \frac{\rhomg_0\,2J}{L_0\,T\exp[3\TC / ((J+1)\,T)]}\,.
\ee

For $T>\TC$ there is $\langle J^z \rangle_{\mrm{MFA}}=0$  and hence
\be\label{51b}
\rhomg = \rhomg_0J(J+1),
\qquad
\Wmag= \frac{\rhomg_0J(J+1)}{L_0\,T}.
\ee

\section{\label{sec:E-P Scattering}Electron--phonon scattering contribution\\ \protect{\hspace*{2em}} to the thermoelectric power}

We consider the phonon system in Debye approximation and the electron--phonon interaction in the deformation potential approximation.
For the phonon system thermodynamical equilibrium is assumed, so no phonon drag processes are considered.
The transition probability per unit time for the free electron normal (i.e. not \textit{Umklapp}) scattering
from the state $\bk$ to the state $\bk'=\bk+\bq$\,,
by phonon of the wave vector $\bq$ and the energy $\hbar\omega_\bq$\,, is:
\be\label{11}
C^{\mrm{ph}}(\bk,\bk\!+\!\bq) = \frac{2\pi}{\hbar}c^\mrm{ph}(q)
\left[\frac{\delta(\vep_{\bk\!+\!\bq}-\vep_\bk-\hbar\omega_\bq)}{\exp[\hbar\omega_\bq/\kBT]-1}
    + \frac{\delta(\vep_{\bk\!+\!\bq}-\vep_\bk+\hbar\omega_\bq)}{1 - \exp[-\hbar\omega_\bq/\kBT]}\right],
\ee
see, e.g., Ref.\,[\onlinecite{zim}], Eq.\,(9.5.6).
For the considered model $\hbar\omega_\bq = \hbar\,q\,v_s$\,,
where $v_s$ denotes the sound velocity averaged over the directions in a~crystal,
and $0\leq q\leq\qD$\,, where $\qD$ is the Debye radius.
The scattering amplitude $c^\mrm{ph}(q)$  has the form\cite{zim}
\be\label{13}
c^\mrm{ph}(q) =  \frac{2 C^2 q}{9 N q_c}\,,
\ee
where $C = 2\vepF/3$ is the interaction energy, $N$ -- the number of ions
(under the assumption of a~one ion of the mass $M$ per a~primitive cell),  $q_c = M v_s / \hbar$.
The first component of the sum (\ref{11}) describes the scattering processes corresponding to the phonon absorption,
and the second one to its emission.

For the electron--phonon scattering, unlike in the case of the electron-wave-vector independent magnetic scattering described in the previous Section, the relaxation-time solution of the Boltzmann equation exists only in the temperatures much greater than the Debye temperature, $\TD$\,.
Thus, for the calculation of the electron--phonon scattering contribution in the transport coefficients the variational method proved to be very useful.

The thermoelectric power $\cSph(T)$ according to the scattering probability  $\Cph(\bk,\bk\!+\!\bq)$ and the formula (\ref{3}) was obtained by Kohler\cite{koh}, and the same form of $\cSph(T)$ was derived by Ziman.\cite{zim}
In the first subsection below (and in Appendix~A) we perform the detailed calculation of the scattering matrix elements $\Pph_{ij}$ (\ref{4}), applying slightly different method than Ziman, but obtaining the same results.

Our way of calculation is similar but simpler than that used by the authors of Ref.\,[\onlinecite{aus2}].
In Appendix~B, we indicate the crucial points in their calculations which led them to the form of the electron--phonon scattering matrix elements differing from those of Kohler (Ziman) and ours, and, in consequence, to the spectacular effects in the behavior of the thermoelectric power.
The same applies to the calculations of the scattering matrix elements in Refs\,[\onlinecite{aus, aus1}], although they have been performed in a~different way than in Ref.\,[\onlinecite{aus2}].
The results for $\Sph(T)$ in each of the papers Refs\,[\onlinecite{aus,aus1,aus2}] are qualitatively the same as it is illustrated in Fig.\,3 of Ref.\,[\onlinecite{aus2}].
For this reason, we refer only to the results of the last paper.

\subsection{The electron--phonon scattering matrix elements} 

Because of the form of $\Cph(\bk,\bk\!+\!\bq)$ it is convenient and natural to
express and calculate the scattering matrix elements $\Pph_{ij}$ (\ref{2}) as the integrals with respect to $\bk$ and $\bq$
($\bq = \bk'-\bk$).
Correspondingly, the functions $u_{ij}(\bk,\bk')$ with the base (\ref{1}) have the form
\be\label{14} \begin{split}
   u_{11} =&\ (\bq\cdot\bu)^2,                                        \\
   u_{12} =&\ (\bq\cdot\bu)^2\,(\vepkq - \zeta(T)) +
         (\bk\cdot\bu) (\bq\cdot\bu) (\vepkq - \vepk),              \\
   u_{22} =&\ (\bq\cdot\bu)^2\,(\vepkq - \zeta(T))^2 + (\bk\cdot\bu)^2 (\vepkq - \vepk)^2 \\
           & +\ 2(\bk\cdot\bu) (\bq\cdot\bu) (\vepkq - \vepk) (\vepkq - \zeta(T)).
\end{split}\ee

Then, as appropriate for a~cubic symmetry, we average (\ref{14}) with respect to all field directions,
$\bu$, $\big( (\bq\cdot\bu)^2 \to q^2/3$, $(\bk\cdot\bu)^2 \to k^2/3$,
$(\bq\cdot\bu)(\bk\cdot\bu) \to (\bk\cdot\bq)/3$\big),
and substitute, by virtue of the energy conservation law,
\begin{displaymath}
\vepkq - \vepk = \pm\zq   \qquad  \bk\cdot\bq = \frac12\, q^2 \left(\frac{\pm\zq}{\vepq} - 1\right)\,,
\end{displaymath}
where
$\zq = \hbar\,q\,v_{s}$,
$\vepq = \hbar^2\,q^2/2m$ is the energy of the electron of the wave vector $\bq$;
the upper sign refers to the absorption and the lower one to the emission.
After representing
$k^2 = q^2 (\eta / \vepq + \zeta(T) / \vepq)$
and denoting
$\eta = \vepk - \zeta(T)$
we obtain
\be\label{15} 
\begin{split}
   u_{11}^\pm =&\ \frac13\, q^2\,,                                                               \\
   u_{12}^\pm =&\ \frac13\, q^2 \left[ \eta \pm \frac{\zq}2 + \frac{\zq^2}{(2\vepq)} \right]\,,   \\
   u_{22}^\pm =&\ \frac13\, q^2
\left[\eta^2 + \frac{2\zq^2\,\eta}{\vepq} \pm \zq\eta + \frac{\zq^2\zeta(T)}{\vepq} \pm \frac{\zq^3}{\vepq} \right]\,.
\end{split}\ee

The scattering matrix elements  $\Pph_{ij}$ can be then written in the  way
\be\label{17} 
\begin{split}
       \Pph_{ij} =&\ {\Pph_{ij}}^+  + {\Pph_{ij}}^-,  \\
   {\Pph_{ij}}^\pm =&\ \frac{V}{\kBT} \int\!\di\bq\, q^2 c^\mrm{ph}(q)\, H_{ij}^\pm(q,\zq),
\end{split}\ee
where
\be\label{18} 
\begin{split}
  H_{11}^\pm(q,\zq) =&\ F_0^\pm\,,                                                          \\
  H_{12}^\pm(q,\zq) =&\ F_1^\pm \pm \frac{\zq}2\, F_0^\pm + \frac{\zq^2}{2\vepq} F_0^\pm\,, \\
  H_{22}^\pm(q,\zq) =&\ F_2^\pm \pm \zq F_1^\pm + 2\frac{\zq^2}{\vepq} F_1^\pm
 + \frac{\zeta(T)\zq^2}{\vepq} F_0^\pm \pm \frac{\zq^3}{\vepq} F_0^\pm\,,
\end{split}\ee
and for $n=0,1,2$:
\be\label{19} 
 F_n^\pm \equiv F_n^\pm (\zq,T) =
 \pm \int\!\di\bk\,(\vepk-\zeta(T))^n\,\delta(\vepkq-\vepk\mp\zq)
 \frac{f^0(\vepk)(1\!-\!f^0(\vepk\pm\zq))}{\exp[\pm z_q/\kBT]-1}.
\ee

After the transformations made in Appendix A  with the use of the approximation of the electron gas strong degeneration the functions (\ref{19}) have the form
\be\label{20} 
F_n^\pm(\zq,T)  =  \frac{\varsigma}{q}\,\, \theta(q_{\max}\!-\!q)\,\, \theta(q)\,\, (\kBT)^{n+1}\,\, I_n(\zq),
\ee
where $\varsigma = 2\pi\,m^2 / (\hbar^4)$, $\theta(x)$ is the Heaviside step function, and 
\ba\label{21} 
I_n^\pm(\zq)  &\equiv&  \pm\!\int\limits_{-\infty}^{\infty}\!\di{y}\,\,y^n f^{\pm}(y,z_q),                 \nonumber\\
f^{\pm}(y,z_q) &=& \frac{(\exp[\pm z_q/\kBT]-1)^{-1}}{(\exp[y]+1)(1+\exp[-(y\pm\zq/\kBT)])}.
\ea

As the immediate consequence of the symmetry property
$f^+(y,z_q) = f^-(-y,z_q)$ of the integrands in $I_n^\pm(\zq)$,
we obtain the following symmetry properties for  $F_n^\pm(\zq,T)$
\ba\label{211}
F_0^+ &=&   F_0^-\,,                                      \nonumber\\
F_1^+ &=& - F_1^-\,,                                  \nonumber\\
F_2^+ &=&   F_2^-\,.
\ea

Accounting (\ref{211}) in (\ref{18}) we reach $H_{ij}^+ (q,\zq) = H_{ij}^- (q,\zq)$ and from (\ref{17}) the conclusion that\\
\hspace*{1em}\textbf{(i)}\,~the contribution to the scattering from the absorption and from the emission must be equal
\be\label{211a}
{\Pph_{ij}}^+  =  {\Pph_{ij}}^-.
\ee

After substitution  $F_n^\pm(\zq,T)$ (\ref{20}) with $I_n^\pm(\zq)$ (\ref{31a}) to (\ref{18}) we get
\be\label{212}\begin{split}
  H_{11}^\pm(q,\zq) =&\ F_0^\pm\,,                                                 \\
  H_{12}^\pm(q,\zq) =&\ \frac{\zq^2}{2\vepq} F_0^\pm                          \\
  H_{22}^\pm(q,\zq) =&\ \left[ \frac{\pi^2}3 (\kBT)^2 + \left(\frac{\vepF}{\vepq}-\frac16\right)\zq^2 \right] F_0^\pm\,,
\end{split},\ee
and we can reach the subsequent conclusion that\\
\hspace*{1em}\textbf{(ii)}\,~the powers $n$ of $q^n$ in all the components in the integrand in $\Pph_{ij}$ (\ref{17}) originating from (\ref{212}) must have the same parity.

The final form of $\Pph_{ij} = 2{\Pph_{ij}}^+$,  (\ref{17}), results from the trivial integration with respect to directions of $\bq$ and using the Debye integrals\cite{zim}
\begin{displaymath}
\int\limits_0^{x} \di z\,\frac{z^n}{\sinh^2(z/2)}  \equiv   4 \cJ_n(x)\,.
\end{displaymath}
After changing the integral variable
$z = \hbar\,qv_{s}/\kBT$, and taking  $\qD = \kB\TD / \hbar\,v_{s}$, one gets
\be\label{26} 
\begin{split}
  \Pph_{11}  =&\ 2\Pph_0 \left(\frac{T}{\TD}\right)^5 \cJ_5 \left(\frac{\TD}T \right)\,,     \\
  \Pph_{12}  =&\ \frac{\veps}2 \Pph_{11}\,,                                    \\
  \Pph_{22}  =&\ 2\Pph_0(\kBT)^2 \left(\frac{T}{\TD}\right)^5
\left[\left(\frac{\pi^2}3 + \frac{\veps\vepF}{(\kBT)^2}\right) \cJ_5 \left(\frac{\TD}T\right)
- \frac16 \cJ_7 \left(\frac{\TD}T\right) \right]\,,
\end{split}
\ee
where
, and
$\veps = 2mv_s^2$ is the energy of the electron of the wave vector $q_s = 2mv_s / \hbar$.

The above result for $\Pph_{ij}$ is the same as $P_{ij}$ derived in Ch.\,IX of Ref.\,[\onlinecite{zim}].
It also corresponds to the result presented in Ref.\,[\onlinecite{koh}],  Eq.(18), after using the equivalence
\be\label{31} 
\frac{\kBT}{\vepF} \,\, \frac{\kBT}{\veps} \equiv \left(\frac{T}{\TD} \right)^2\, \frac1{n_s}\,,
\ee
taken for $n_s = (1/2)^{2/3}$. 
The parameter $n_s$ is related to the electron gas density $n_a$\,, $n_s = (n_a / 2)^{2/3}$, and for metals  $n_s \geq (1/2)^{2/3}$.

Notice that the $n$ of $\cJ_n(u)$, occurring in all $\Pph_{ij}$ (\ref{26}), has the same parity according to the conclusion (ii) above.

\subsection{\label{sec:Inelastic Contribution}%
Electron--phonon scattering contribution to the thermoelectric power
}

The electron--phonon contribution to the thermoelectric power $\cSph(T)$ results from substitution in (\ref{3}) $\Pph_{ij}$  in place of $P_{ij}$\,.
Considering it as the function of the reduced temperature $t = T / \TD$ and applying the identity (\ref{31}) we get after some algebra
\ba\label{213}
\cSph(t)  =   - \frac{\pi^2}3\, \frac{\kB}{e}\,  \frac{\kB\TD}{\vepF} \,t\,\,R^{\mathrm{ph}}(t)                        \nonumber\\
R^{\mathrm{ph}}(t)  =  \frac{\pi^2\,t^2 - \cJ_7\,t^2/(4\cJ_5) + n_s}{\pi^2\,t^2/3 - \cJ_7 \, t^2 / (6\cJ_5) + n_s},
\ea
where $\cJ_n\equiv\cJ_n(1/t)$.

Calculating $\cSph(t)$ (\ref{213}), which depends on $n_s$\,, the Fermi energy $\vepF$ and the Debye temperature $\TD$, we fixed the last two parameters, similarly as in Ref.\,[\onlinecite{aus2}], assuming the same values $\vepF = 1.5$\,eV and $\TD = 200$\,K, as were fixed therein.
The values of $n_s$ corresponding to the values of $\vep_s$\, borrowed from Ref.\,[\onlinecite{aus2}], and the other material constants, are presented in TABLE~I, providing the corrected (with respect to the values in the Table~I\,\cite{aus2}\,)
correspondence between $\veps$, $v_s$ and $\qD$\,.

Irrespective of the numerical analysis of $\cSph(t)$, which we present as the graphs in Fig.\,1, some general conclusions can be obtained from the very analysis of the above formula.
The behavior of $\cSph(t)$, illustrated in Fig.\,1, is determined by the properties of the function $R^{\mathrm{ph}}(t)$:\\

(I) $ 1<R^{\mathrm{ph}}(t)< 3$;\\[-1.2ex]

(II) $\lim_{t\rightarrow 0}R^{\mathrm{ph}}(t)=1$, \quad   $\lim_{t\rightarrow \infty}R^{\mathrm{ph}}(t)=3$;\\[-1.2ex]

(III) $\displaystyle \frac{ \pt R^{\mathrm{ph}}(t,n_s) } { \pt n_s} < 0$.\\[-1.2ex]

(I)--(III)  can be derived directly from the formula for
$R^{\mathrm{ph}}(t)$ (\ref{213})
if one takes into account the metallic values of
$n_s$ ($n_s \geq (1/2)^{2/3}$)
and uses the estimation
$\cJ_7(1/t)/\cJ_5(1/t)< (1/t)^2$ for $t>2^{1/2}/4\pi$
and the  approximation
$\cJ_7(1/t)/\cJ_5(1/t)\simeq 7!\zeta(7)/5!\zeta(5)$ for $t\leq 2^{1/2}/4\pi$.
The last follows from $\cJ_n(x\gg 1)\simeq n!\zeta(n)$ -- the approximation of the Debye integrals by the zeta-Riemann function $\zeta(n)$ for $x\gg 1$ ($\zeta(7)/\zeta(5)\simeq 1$).

With the use of (I)--(III) we get the conclusion that $\cSph(t)$ is bounded by the linear functions
\be\label{214}
-\pi^2\, \frac{\kB}e \; \frac{\kB\TD}{\vepF}\,t < \cSph(t) < - \frac{\pi^2}3\, \frac{\kB}{e}\, \frac{\kB\TD}{\vepF}\,t\,,
\ee
being its asymptotes, correspondingly  for $t\rightarrow \infty$ and for $t\rightarrow 0$.

The subsequent conclusion, which  we get  from (\ref{214}) and (\ref{50h}) is that for every $t>0$
\be\label{214a}
    1/3  <  \frac{ \cS^{mag}(t)}{\cS^{ph}(t)} < 1\,,
\ee
where $\Smag(t)= -\pi^2\kB\TD\,t / (3\vepF)$.

\begin{table} \centering
\caption{%
The parameter $n_s = (n_a/2)^{2/3}$, related to the electron gas density $n_a$\,,
the sound velocity $v_s$ and the Debye radius $\qD$\,, corresponding to values of $\vep_s$\,.
}
\vspace{3ex}
\begin{tabular}{cccc}
\hline\hline 
 $\vep_s$\,[K] & $n_s$ & $v_s$\,[m/s] & $\qD$\,[nm$^{-1}$] \\
\hline 
     1.5  &  0.652  &  4770   &   5.49 \\
     3.0  &  1.30   &  6740   &   3.88 \\
     6.0  &  2.61   &  9540   &   2.75 \\
    12.0  &  5.22   & 13500   &   1.94 \\
    24.0  & 10.43   & 19100   &   1.37 \\
    48.0  & 20.87   & 27000   &   0.97 \\
\hline\hline 
\end{tabular}
\end{table}

\begin{figure}
\begin{center}
\includegraphics[angle=270,width=0.95\columnwidth]{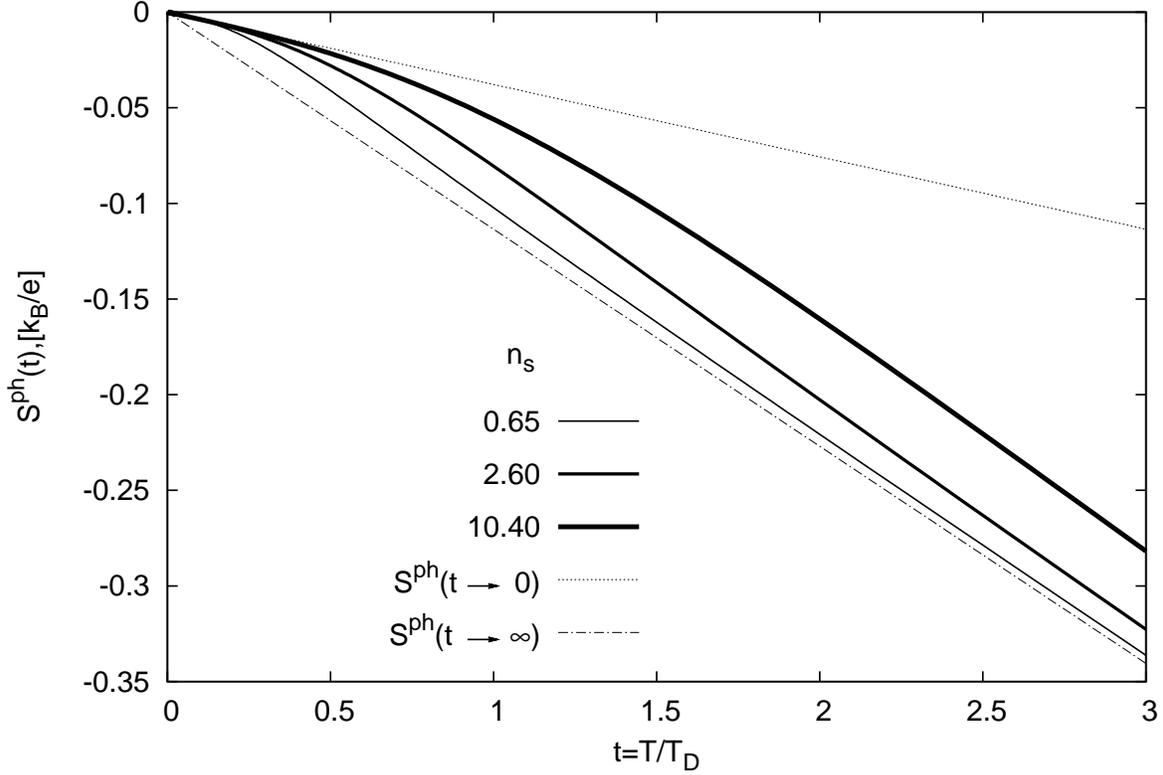}
\caption[Fig.1]{%
The electron--phonon scattering contribution to the thermoelectric power $\cSph(t)$ (\ref{213}) labeled by the parameter $n_s$\,, TABLE~I.}
\end{center}\end{figure}

The limitations of $\cSph(t)$ and its asymptotic behavior is seen from the graphs in Fig.\,1.

Additionally, on the basis of (III) we can conclude that $\cSph(t,n_s)$ (\ref{213}) increases its values for the increasing values of $n_s$\,.
This conclusion is also illustrated by the graphs in Fig.\,1.

Our results can be compared with those presented in Ref.\,[\onlinecite{aus2}], which depend on $\veps$ (see Fig.\,2 therein), if one takes into account the relation between $n_s$ and $\veps$\,, see~TABLE~I.

Note at the end of this section, that after substituting  $\ovl{\Pph_{11}}$,  $\ovl{\Pph_{22}}$ (\ref{26}) and  $J_1$\,, $U_2$ (\ref{500k}) into (\ref{4}), one gets the standard results for the electron--phonon part of the thermal $\Wph(t)$ and the electrical  $\rhoph(t)$ resistivity, Refs\,[\onlinecite{koh,koh1,zim}].
\ba\label{215}
  \Wph(t)  &=&  \frac{\rhoph(t)}{L_0\,\TD}\,\frac1{t} \,
  \left[1 + \frac3{\pi^2}\,  \frac{n_s}{t^2} - \frac1{2\pi^2}\, \frac{\cJ_7(1/t)}{\cJ_5(1/t)} \right],      \nonumber\\
   \rhoph(t) &=& \rhoph_0 \, t^5\cJ_5(1/t)\,,
\ea
where  $\rhoph_0 = 3\pi m^2 C^2 \qD^5 / (16 e^2 N Mv_s \kF^6) $.

For $t\ll 1$, because of $\cJ_{5}(1/t)=5!\zeta(5)$, there is
\be\label{215a}
\rhoph  \sim  \rhoph_0\,t^5,
\qquad\qquad
\Wph  \sim  \frac{\rhoph}{L_0\,\TD}\, \frac3{\pi^2}\,n_s\,t^2\,.
\ee

For $t\gg 1$ one can approximate  $\cJ_{5}(1/t)\simeq 1/(4t^4)$ and hence
\be\label{215b}
\rhoph  \sim  \rhoph_0\,t,
\qquad\qquad
\Wph  \sim  \frac{\rhoph_0}{L_0\,\TD}.
\ee

\section{The total thermoelectric power from~the~electron--\\
\protect{\hspace*{2em}} --phonon and~the~electron--localized~spin scattering}

In order to examine  the  dependence  of the total thermoelectric power $\cS(t)$ on the reduced temperature $t=T/\TD$ we use Kohler rule (\ref{8}).
Some general conclusions about $\cS(t)$ can be drawn directly from this rule when we account the results of Sections III and IV.B.  Writing (\ref{8}) in the form
\be\label{216}
\cS(t)  =  \cSph(t)\, \frac{1 + [\cSmg(t)/\cSph(t)][ \,W^{mag}(t)/W^{ph}(t)]}{1+ \Wmag(t)/\Wph(t)}\,,
\ee
and applying (\ref{214})--(\ref{214a}) one can easily find that $\cSph(t) < \cS(t)$.
Similarly,\, 
by the mutual exchange of indices `ph' and `mag' in (\ref{216}), it can be shown that $\cS(t) < \cSmg(t)$.
Summarizing, we get for $\cS(t)$
\be\label{217}
 -\pi^2\, \frac{\kB}e \; \frac{\kB\TD}{\vepF}\,t < \cS(t) < -\frac{\pi^2}3 \frac{\kB}{e}\, \frac{\kB\TD}{\vepF}\,t\,
\ee
the same limiting conditions, as these for $\cSph(t)$, (\ref{214}).
Additionally, for $t\ll 1$ there is
$\Wmag/\Wph\sim t^{-2}\exp[ -3\TC / (\TD(J+1)]$   (see (\ref{51a}), (\ref{215a}))
and for $t\gg 1$ correspondingly
$\Wmag/\Wph \sim \gamma\,J(J+1)/t$,
where
$\gamma=\rhomg_0/\rhoph_0$ (see (\ref{51b}) and (\ref{215b})).
Applying this in (\ref{216}) and including (\ref{214a}), we obtain that $\cS(t)$ shows the same asymptotic behavior in low and high temperatures as~$\cSph(t)$.

The properties of $\cS(t)$ discussed above  can be seen from the results of the numerical calculations presented as the graphs in Fig.\,2.
We have obtained them for the set of parameters used in Ref.\,[\onlinecite{aus}] as corresponding to GdAl$_2$: $\TD = 289$\,K, $\TC = 180$\,K, $\kB\TD / \vepF = 0.0025$,  $\gamma=0.033$, $n_s = 4.07$, $J = 3.5$ (see Table\,2 therein), and additionally for various values of $n_s$\,, $J$ and $\gamma$.

\begin{figure}
\begin{center}
\includegraphics[angle=270,width=0.95\columnwidth]{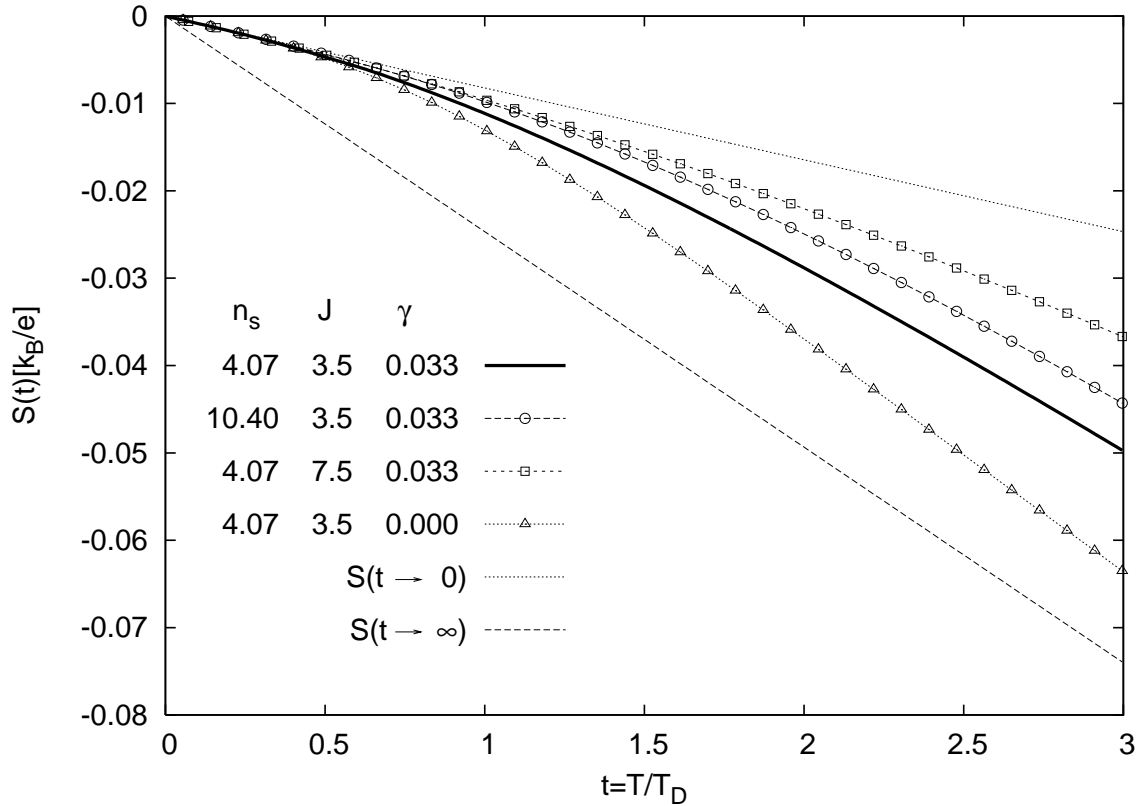}
\caption[Fig.2]{%
The total thermoelectric power $\cS(t)$ (\ref{218}) for different values of the parameters $n_s$\,, $J$ and $\gamma=\rhomg_0 / \rhoph_0$ and for the fixed  $\TD = 289$\,K, $\TC = 180$\,K, $\kB\TD / \vepF = 0.0025$.
The continuous, thick line corresponds to the data for GdAl$_2$ according to Table\,2 in Ref.\,(\onlinecite{aus}).}
\end{center}\end{figure}

The  dependence of $\cS(t,p)$  on the free parameter $p=n_s, \gamma, J$ can be examined in an analytical way, similarly as for $\cSph(t)$, after writing the formula (\ref{216}) in the form
\ba\label{218}
\cS(t) &=& -\frac{\pi^2}3 \, \frac{\kB}e \, \frac{\kB\TD}{\vepF} \,\,t\,R(t)                            \nonumber\\
 R(t)  &=&  \frac{3+(3/\pi^2)n_s/t^2\!-\!(3/4\pi^2)\cJ_7/\cJ_5+\rhomg(t)/\rhoph(t)}%
{1+ (3/\pi^2)n_s/t^2\!-\!(1/2\pi^2)\cJ_7/\cJ_5+\rhomg(t)/\rhoph(t)}\,,
\ea
resulting from the substitution in (\ref{216}) $W^{\alpha}(t)$ (\ref{50m}), (\ref{215}) and  $\cS^{\alpha}(t)$ (\ref{50h}), (\ref{213}), ($\alpha=$\,mag,\,ph).

It can be shown, by repeating the reasoning conducted previously in relation to $R^{\mathrm{ph}}(t,n_s)$ that $\pt R(t,p)/\pt p\, < 0$  for  $p = n_s,\gamma$ and $t > 0$ ($\rhomg(t)) / \rhoph(t)$ is linear with respect to $\gamma$).
The similar justification for $\pt R(t,J)/\pt J\, < 0$ can be performed under the condition $t > \TC/\TD$, where $\rhomg(t)$ (\ref{51b}) depends sufficiently simply on $J$.

We can conclude that $\cS(t,p)$ (\ref{218}) is the increasing function of $p=n_s, \gamma, J$\,  in the appropriate range of temperature (in the case of  $J$  for $t>\TC/\TD$).
This property is illustrated in Fig.2.

\section{\label{sec:Summary}Summary and conclusions}

We have calculated the contributions to the thermoelectric power of the f-electron metals from the s--f scattering,  $\cSmg(T)$, and from the electron--phonon scattering, $\cSph(T)$, applying the same standard approach as the authors of Refs\,[\onlinecite{aus,aus1,aus2}] (Born approximation for the scattering and Kohler (Ziman) variational formula for the thermoelectric power).
For the total thermoelectric power $\cS(T)$ we have used Kohler rule,\cite{koh1} equivalently to using the variational formula and the Matthiessen rule for the scattering probabilities in Ref.\,[\onlinecite{aus}].

In the case of the s--f scattering, basing on the symmetry of the scattering probability in the molecular field approximation for the f-electron system, we have shown the equivalence of the variational formula and the Mott formula (with the relaxation time dependent on the electron energy as $\tau\sim \vep^{-1/2}$).
This gave us the linear dependence of $\Smag(T)$ with the negative sign.

For the electron--phonon scattering we performed the detailed calculation of the scattering matrix elements by an equivalent method to that applied in Ref.\,[\onlinecite{ aus2}] and, unlike them, we have got the results, obtained previously by Kohler and Ziman.
We have shown that these matrix elements satisfy some symmetry conditions, which we have formulated as the conclusions (i)--(ii) in Section~IV.A.

Analyzing the variational formula for $\cSph(T)$ with these scattering matrix elements, we were able to show that  $\cSph(T)$ has the constant and negative sign, being bounded by its high-temperature and low-temperature asymptotes.
The same concerns $\cS(T)$, as we have shown with the use of Kohler rule.

From the results of our numerical calculations,  presented in Fig.\,1, the weak nonlinearity in the intermediate temperatures depends, for the constant Fermi energy and the Debye temperature, on the electron gas density characterized by the parameter $n_s$\,.
Moreover, as we could state analyzing the formula for $\cSph(T)$, it is the increasing function of $n_s$\,, what is also seen from Fig.\,1.

The same nearly linear behavior in low and high temperatures and weakly nonlinear in the intermediate temperatures is exhibited by the graphs of the total thermoelectric power $\cS(T)$ in Fig.\,2, obtained similarly as $\cS^{ph}(T)$ for the fixed values of the Fermi energy and  the Debye temperature, and, additionally, for the constant Curie temperature.
The graphs in Fig.\,2 illustrate also the property of $\cS(T)$, which we have derived analytically, that it is the increasing function of the parameters:  $n_s$\,,  the quantum number $J$ and the parameter $\gamma$, characterizing the relative contribution of the s--f scattering.
For the rising values of the last two parameters $\cS(T)$, as it is seen from Fig.\,2, approaches $\Smag(T)$.

Analyzing the way of calculation of the electron--phonon scattering matrix elements in Ref.\,[\onlinecite{aus2}], we have found that some terms were omitted in the integrands, and also the contribution to the scattering from the emission processes was completely omitted.
In consequence of these inaccuracies, the scattering matrix elements derived in Ref.\,[\onlinecite{aus2}] do not fulfill the symmetry conditions which should be fulfilled.
This can be seen by comparing (\ref{53}) and (\ref{55}) with (i)--(ii) in Section~IV.A.

The same  concerns the scattering matrix elements presented in Refs\,[\onlinecite{aus,aus1}], as we could state performing calculations by the way applied there -- the slightly different one than that applied in Ref.\,[\onlinecite{aus2}].
We can conclude that the sign reversal and the maxima of $\cSph(T)$ obtained in Refs\,[\onlinecite{aus,aus1,aus2}] and these of $\cS(T)$ in Ref.\,[\onlinecite{aus}] have their origin in the flaws in the scattering matrix elements calculations, which we have described above. The direct meaning of this conclusion is that the interpretation of the anomalous behavior of TEP can not be based on these simple models which have been applied in Refs\,[\onlinecite{aus,aus1,aus2}]. However, it also restores the question about an adequate models for explanation of the thermoelectric power anomalies in the metallic f-electron systems. This concerns particularly the anomalies in low temperatures which are not explained by dynamically developing theories
of strongly correlated electron systems. In the Introduction we gave only a partial answer to this question.
We have payed attention there on the models described in Refs\,[\onlinecite{tak,kas}], which can  explain the  anomalies in paramagnetic systems with crystal-field (CF) splitting\,\cite{tak} or in the ferromagnets\,\cite{kas}- not taking this splitting into account.

There are  examples of  the f-electron systems   with CF splitting, the thermoelectric power of which exhibits the anomalies in the magnetically ordered phase,   see  Ref\,[\onlinecite{a7,a8,a9,n0}] or
examples  in Ref\,[\onlinecite{fou}].  Finding the appropriate models  is therefore an important issue for further research. In particular, an interesting problem could be the
extension of considerations of Refs\,[\onlinecite{tak,kas}] for the cases of the anomaly in the ordered phase of the f-electron systems with the CF splitting.

\appendix 
\section{} 

Integration in $F_n^\pm(\zq,T)$  (\ref{19})  with respect to the angle
$\THeta = \angle(\bk,\bq)$
can be done by the integration with respect to $\bk$ in the spherical coordinate system,
where $\bq$ is parallel to the polar axis. With the use of the delta-Dirac function properties and changing the variable
$y = (\vepk-\zeta(T)) / \kBT$ in the integration with respect to $k$ we have
\be\label{16b} 
F_n(q,\zq)^\pm = \frac{(\kBT)^{n+1}\,m^2}{\hbar^4q}\;\;
\pm\!\!\!\!\!\!\!\!\int\limits_{\XI(q,\mp\zq,T)/\kBT}^{\infty}\!\!\!\!\!\!\di{y}\;
 \frac{y^n\,(\exp[\pm z_q/\kBT]-1)^{-1}}{(\exp[y]+1)(1+\exp[-(y\pm\zq/\kBT)])}\,,
\ee
where $\XI(q,\mp\zq,T) = (\hbar^2/2m)\left|\zq\,m/\hbar^2\,q\mp q/2\right|^2-\zeta(T)$.

In the approximation of the strong degeneration of the electron gas and for the experimentally accessible temperature range there is
\begin{displaymath}
\frac{-\zeta(T)}{\kBT}  \simeq  -\frac{\vepF}{\kBT}  \simeq  -\infty ,
\end{displaymath}
and $\XI(q,\mp\zq,T)$ can be considered as temperature-independent for that temperature range.
Thus, for $q$  fulfilling the inequality
$\XI(q,\mp\zq) < 0$ ($z_q = \hbar\,q\,v_s$) and the equivalent one
\be\label{17a} 
\left| \frac{q}2 \mp q_s \right|^2  -  \kF^2\,  <  0,
\ee
where $q_s = mv_s / \hbar$, the lower limit of the integral can be replaced by $-\infty$, for the sufficiently low $T$.
Comparing the typical for metals value of the sound velocity,
$v_s = \hbar q_s / {m} < 6 \cdot 10^3$\,m/s,
with the value of the Fermi velocity,
$\vF = \hbar \kF / {m} \simeq 4\cdot10^5$\,m/s,
one gets
$q_s / \kF \simeq 10^{-2}$.
Because for metals there is $\qD \leq 2^{1/3}\kF$
the range $0 \div \qD$ can be accepted as the solution of the inequality (\ref{17}), both for the phonon absorption and emission.
It justifies the approximation $\XI(q,\mp \zq, T)/\kBT  \simeq  - \infty$  in (\ref{16b}) which now can be represented in the form (\ref{20})--(\ref{21}).

The integrals $I_n^\pm(\zq)$ (\ref{21}) can be calculated with the aid of the formula\cite{zim}
\be\label{22a} 
\begin{split}
 \int\limits_{-\infty}^{\infty}\!\di{y}\; \frac{\cF(y)}{(\exp[y]+1)(1+\exp[-(y+z)])} =&\
 \int\limits_{-\infty}^{\infty}\!\di{y}\; \left[ \frac{\cG(y)-\cG(y-z)}{(1-\exp[-z])} \right]
 \left( \frac{-\pt f^0(y)}{\pt y} \right), \\
\cG(y) =&\ \int\limits_0^y\!\di{y'}\,\cF(y')\,,
\end{split}\ee
and next with the use of the Sommerfeld expansion
\be\label{22b}
\int\limits_{-\infty}^{\infty}\!\di{y}\, \left( -\frac{\pt f^0(y)}{\pt y} \right)\,\cH(y)
= \cH(0) + \frac{\pi^2}6\, \left.\frac{\pt^2\,\cH(y)}{\pt y^2}\right|_{y\,=\,0} + \cdots\,.
\ee

When $\cF(y)$ in the integrand (\ref{22a}) has the form of polynomial, like in $I_n^{\pm}(\zq)$, the first term of the above expansion is the exact value of the integral, what gives
\be\label{31a} 
\begin{split}
   I_0^\pm(\zq) =&\ \frac{4\zq/\kBT}{\sinh^2[\zq/2\kBT]-1}\,,                                   \\
   I_1^\pm(\zq) =&\ \pm\frac{\zq\,I_0^\pm(\zq)}{2\kBT}\,,                                       \\
   I_2^\pm(\zq) =&\ \left[\frac{\pi^2}3 + \frac{\zq^2}{3(\kBT)^2} \right]\, I_0^\pm(\zq)\,.
\end{split}\ee


\section{} 

Comparing $u_{ij}(\bk,\bk+\bq)$ (\ref{14}) with the equation (3.8) of Ref.\,\onlinecite{aus2}, one can see that the component
$(\bq\cdot\bu)(\bk\cdot\bu)$ has been omitted there.
Since the authors of Ref.\,\onlinecite{aus2} had used this incomplete form $\widetilde{u_{ij}}(\bk,\bk+\bq)$
as non-averaged over field directions, they had to take into account the space directions of the vectors $\bq$ and $\bk$
when calculating the integrals in the scattering matrix elements.
This made the integration more complex than in our calculations presented in Section IV.A. \
Despite this, our method of calculation, although simpler, is equivalent to theirs.

Performing on  $\widetilde{u_{ij}}$  the same transformations which led from (\ref{14}) to (\ref{15}) one gets
\be\label{16a} 
\begin{split}
\widetilde{u_{11}^\pm} =&\ \frac13\, q^2\,,                                                  \\
\widetilde{u_{12}^\pm} =&\ \frac13\, q^2(\eta\pm\zq)\,,                                      \\
\widetilde{u_{22}^\pm} =&\ \frac13\, q^2\,
\left[ \eta^2 \pm 2\zq\eta + \zq^2 + \frac{\zq^2\eta}{\vepq} + \frac{\zq^2\zeta(T)}{\vepq} \right].
\end{split}\ee

Repeating with (\ref{16a}) in place of (\ref{15}) all the calculations made in Section IV.A and Appendix\,A one obtains the form of (\ref{212}) which we denote as $\widetilde{H_{ij}^{\pm}}(q,\zq)$
\be\label{53} 
\begin{split}
\widetilde{H_{11}^\pm}(q,\zq) =&\ F_0^\pm\,,                                                \\
\widetilde{H_{12}^\pm}(q,\zq) =&\ \pm \frac{\zq}2 F_0^\pm                                   \\
\widetilde{H_{22}^\pm}(q,\zq) =&\ \left[ \frac{\pi^2}3(\kBT)^2
  + \left(\frac{\vepF}{\vepq} + \frac13\right)\zq^2 \mp \frac{\zq^3}{2\vepq} \right] F_0^\pm\,.
\end{split}\ee

Substitution  (\ref{53}) in (\ref{17}), in place of (\ref{18}), leads to the non-equivalence of the contributions to the scattering matrix elements from the absorption and the emission processes $\widetilde{P^{ph+}_{ij}}\neq\widetilde{P^{ph-}_{ij}}$.
However the authors of Ref.\,(\,\onlinecite{aus2}) have omitted the emission processes in their calculations.
It can be easily verified by the substitution
$\widetilde{H_{ij}^{+}}(q,\zq)$ (\ref{53}) in ${\Pph}^+_{ij}$ (\ref{17}) in place of $H_{ij}^{+}(q,\zq)$.
Performing next all the transformations described in Section IV.A, which led to the final form of $\Pph_{ij}$ (\ref{26})
one gets the scattering matrix elements, the averaged  $\ovl{\PDA_{ij}} = \PDA_{ij}/(\Pph_0(\kBT)^{i+j-2}\,t^5)$ form of which
\be\label{55} 
\begin{split}
\ovl{\PDA_{11}}  =&\  \cJ_5(1/t)\,, \\
\ovl{\PDA_{12}}  =&\ \frac{1}{2} \cJ_6(1/t)\,, \\
\ovl{\PDA_{22}}  =&\ \left[\left( \frac{\pi^2}3 + \frac{\veps\,\vepF}{(\kBT)^2}  \right) \cJ_5(1/t) -\frac{\veps}{2\kBT}\cJ_6(1/t)+ \frac13 \cJ_7(1/t) \right]\,,
\end{split}\ee
is the same as those in the equations (4.2a)--(4.2c) in Ref.(\,\onlinecite{aus2}), corrected by the Erratum\cite{aus2}.

\begin{flushleft}

\end{flushleft}

\end{document}